# Electronic Structure and Optical Absorption of Fluorographene

Yufeng Liang and Li Yang

Department of Physics, Washington University in St. Louis, One Brookings Drive, St. Louis, MO 63130, USA

**ABSTRACT**

A first-principles study on the quasiparticles energy and optical absorption spectrum of fluorographene is presented by employing the GW + Bethe-Salpeter Equation (BSE) method with many-electron effects included. The calculated band gap is increased from 3.0 eV to 7.3 eV by the GW approximation. Moreover, the optical absorption spectrum of fluorographene is dominated by enhanced excitonic effects. The prominent absorption peak is dictated by bright resonant excitons around 9.0 eV that exhibit a strong charge transfer character, shedding light on the exciton condensation and relevant optoelectronic applications. At the same time, the lowest-lying exciton at 3.8 eV with a binding energy of 3.5 eV is identified, which gives rise to explanation of the recent ultraviolet photoluminescence experiment.

**INTRODUCTION**

Graphene [1] has attracted intensive attention recently because of its extraordinary physical properties and broad applications. The linear dispersion relation near the Dirac point results in the relativistic nature of electrons [2] and leads to an ultrahigh mobility of charge carriers, rendering graphene a promising candidate for future high speed nano-electronic devices [3]. However, one known obstacle of putting graphene into realistic applications is the absence of a finite band gap that is essential for semiconductor devices, such as building bipolar junction structures. As a result, many efforts have been performed to generate a finite band gap in graphene or its derivatives. Among them, the chemical modification [4-7] has been regarded as a promising candidate because of its low cost and capability for large-scale productions. Through this chemical approach, the adsorbed atoms or chemical groups on graphene surface could strongly affect the band structure of $\pi$-electrons. For example, the attached atoms may form covalence bonds with graphene and transform the $sp^2$-hybridization of graphene into a $sp^3$-hybridization. Consequently the relevant $\pi$ bands crossing at the Dirac point are perturbed, generating a finite band gap.

To date two typical chemically functionalized graphene derivatives have been realized in experiments, i.e., hydrogenated [5] and fluorinated graphene [7]. They exhibit a pronounced band gap under appropriate conditions [5-9]. More importantly, their band gap can be tuned in a wide range (a few eV) [6, 7] by controlling the density of adsorbed atoms. This provides us a precious degree of freedom to tailor the electronic structure of graphene for specific applications [7, 9, 10]. In this work, we are especially interested in fluorinated graphene because of the extremely low electron affinity of fluorine, which is apt to make the fluorinated graphene an *n*-doped semiconductor and hence a better prototype material for electronic devices.

Although there are a number of studies on the electronic structure of fluorinated graphene, very few of them have included many-electron effects that are known to be of importance to decide the electrical and optical properties of reduced dimensional structures. Particularly in terms of the optical response, electron-hole *(e-h)* interactions have been shown to be necessary for a trustable optical absorption spectrum [18, 19]. All these problems motivate us to perform a systematical first-principles study on the electronic structure and optical response of fluorinated graphene with many-electron effects included. This study will provide the fundamental parameters for nanodevice design based on this novel material.

In the following calculations, we focus on the fully fluorinated graphene (fluorographene) in the chair configuration because of the distinct chemical stability of this configuration [7, 8, 11]. We start from the density functional theory (DFT) under the local density approximation (LDA) to obtain the Kohn-Sham particle eigenenergy and wave functions. Then, we apply the GW approximation to calculate the quasiparticle energy and hence give the band gap and band structure. Finally, we solve the BSE to obtain excitonic states and the corresponding optical absorption spectrum with *e-h* interactions included.

**THEORY**

The ground-state electronic structure is obtained by solving the Kohn-Sham equation within LDA [15-16]. We use Troulier-Martins norm-conserving pseudopotentials [17] by employing a plane-wave basis with a 60-Ry energy cutoff. Next, we compute the quasiparticle energy correction by the $G_0W_0$ approximation [18]. The plane-wave energy cutoff of the dielectric function is set to be 8.0-Ry for a full convergence and the static dielectric function is extended to the dynamic case with the generalized plasmon-pole model [18, 21]. A $16 \times 16 \times 1$ k-point sampling is sufficient to calculate the quasiparticle energy as a result of the wide-gap nature of fluorographene [7-11]. To eliminate the artificial interactions inherent in the supercell configuration, we truncate the Coulomb potential between two adjacent fluorographene layers. Finally, the optical absorption spectrum with excitonic effects included is considered by solving the BSE

$$(E_c - E_v)A^s_{vck} + \sum_{v'c'} K^{AA}_{vck,v'c'k}(\Omega_S)A^s_{v'c'k} = \Omega_S A^S_{vck}, \qquad (1)$$

with a static *e-h* interaction approximation [19, 20]. All intricate *e-h* interactions have been embodied in the frequency-dependent kernel $K^{AA}_{vck,v'c'k}(\Omega_S)$ and the *e-h* wave function is represented by the superposition

$$\chi_S(x_e, x_h) = \sum_k \sum_v \sum_c A^S_{vck} \psi_{c,k}(x_e) \psi^*_{v,k}(x_h), \qquad (2)$$

where $A^S_{vck}$ is the exciton amplitude, and $\psi_{ck}(\vec{x}_e)$ and $\psi_{vk}(\vec{x}_h)$ are the electron and hole wave functions, respectively. $k$ is the wave vector of periodic systems. To simulate an absorption spectrum with satisfactory convergence up to 20 eV, we consider 7 valence and 9 conduction bands and adopt a 40 × 40 × 1 uniform k-grid on the first Brillouin zone (BZ).

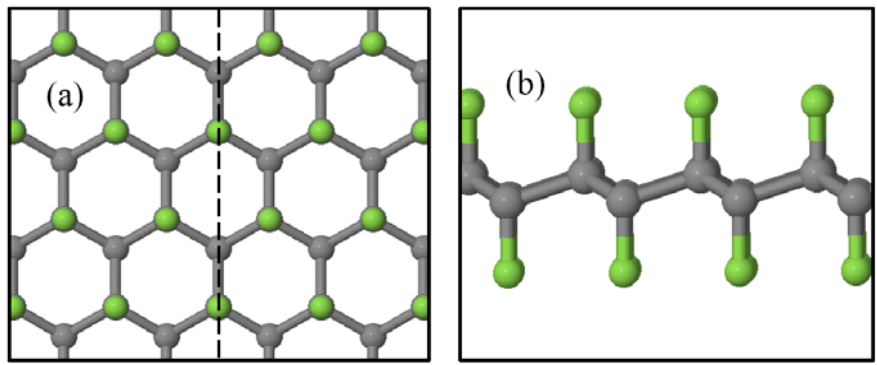

**Figure 1. (a)** Top view of the chair configuration of fluorographene. **(b)** Side view along the armchair direction marked by the dash line in (a). Grey balls represent carbon atoms whereas green balls represent fluorine atoms.

**DISCUSSION**

The structure of fluorographene is fully relaxed according to the force and stress within DFT/LDA, as is shown in **Fig. 1**. Each carbon atom is covalently bonded to a single fluorine atom such that two adjacent C-F bonds are in an antiparallel orientation. This is known as the chair configuration and has been proved to be more energetically favorable than other configurations [10]. Therefore, we only consider this structure in the following calculations.

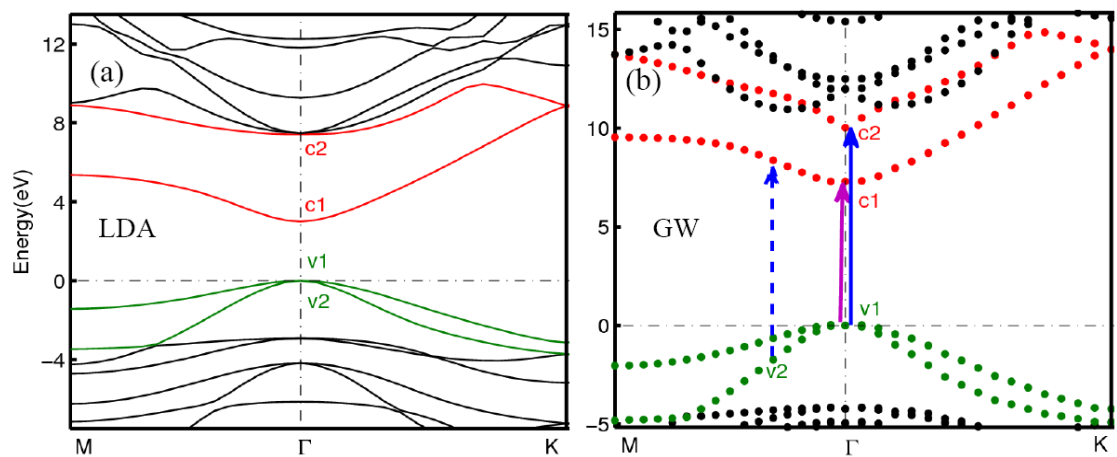

**Figure 2.** LDA **(a)** and GW **(b)** band structure along two high-symmetry directions of fluorographene. The energy reference level has been chosen to be the valence band maximum. The dominant transitions in the optical absorption spectrum are colored with purple or blue and the bands (v1, v2, c1 and c2) involved are colored with red or green.

The LDA band structure of fluorographene is displayed in **Fig. 2(a)**. As expected, the π and π* bands are completely eliminated by the full fluorination. Our DFT calculations show that the strong covalent bonding of the carbon and fluorine atoms causes a large direct band gap (3.0eV) at the Γ point with double degenerated valence bands. In order to understand the electronic structure of fluorographene, we have plotted the charge distribution of a few typical electronic states in **Fig. 3**. For example, the valence states at the Γ point are $sp^3$-hybridization orbits forming carbon-carbon and carbon-fluorine σ bonds as shown in **Fig. 3 (a)**. Especially, a significant electron charge is attracted to fluorine atoms because of their strong electron negativity. On the other hand, the electronic states in conduction bands demonstrate an opposite trend of charge distribution. For example, the lowest (C1) and second lowest (C2) conduction states are mainly localized on the vertical carbon-fluorine bond direction as shown in **Fig. 3 (b) and (c)**. All these wave functions differ from those in graphane in that the electron tends to concentrate in the vicinity of the fluorine atoms.

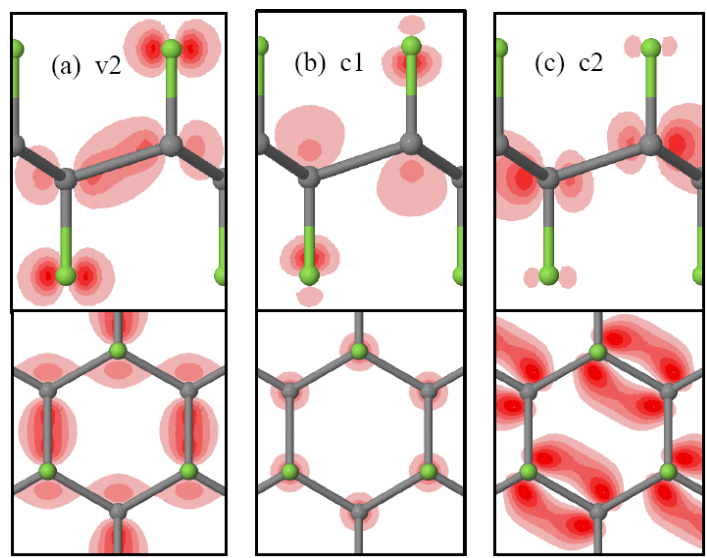

**Figure 3.** Charge distributions of the electronic states v2, c1 and c2 labeled in **Fig. 2**. They are plotted in a plane crossing the dash line in **Fig. 1(a)** and vertical to the fluorographene plane (upper half in each subfigure) and a plane from the top view (lower half), respectively.

The GW-approximation corrected quasiparticle band energies are depicted by the blue circle in **Fig. 2 (b)**. The quasiparticle band gap increases up to 7.3 eV, suggesting an enhanced electron-electron interaction in fluorographene. This is due to the reduced electron screening effects in low-dimensional semiconducting materials. Our GW results are in reasonable agreement with previous studies [11, 13, 14] and once again confirm the wide gap property of fluorographene [7]. Moreover, we notice that the GW correction is significantly higher compared with that of graphane, in which the fundamental band gap is only modified from 3.4 eV to 5.4 eV [12] by many-electron effects. We attribute the larger GW correction in fluorographene to its denser charge distribution near fluorine atoms, given the pronounced electronegativity of fluorine atoms and the fact that self-energy corrections are usually more significant to those electronic states with a higher spatial density.

|  | Γ-K | | Γ-M | |
| --- | --- | --- | --- | --- |
|  | LDA | GW | LDA | GW |
| Light hole | 0.29 | 0.21 | 0.31 | 0.23 |
| Heavy hole | 0.53 | 0.36 | 0.84 | 0.51 |
| Electron | 0.32 | 0.27 | 0.43 | 0.33 |

**Table 1.** Effective mass along Γ-K and Γ-M directions based on LDA and GW calculations. The values are in the unit of free electron mass ($m_e$).

In addition to the band gap correction, self-energy corrections also have marked impacts on the effective mass of free carriers in fluorographene. Because the effective mass is usually a tensor, we have listed calculated values along two typical directions in **Table 1**. I can be seen that the GW corrections usually decrease the effective mass of all light hole, heavy hole and electron at the Γ point. For example, the many-body effect is so prominent that the effective mass of the heavy hole is reduced from 0.53 $m_e$ to 0.36 $m_e$, which is a 32% reduction. Therefore, due to many-electron interactions, fluorographene may exhibits a better mobility of free carriers and than LDA predictions, which is of importance for high-speed devices.

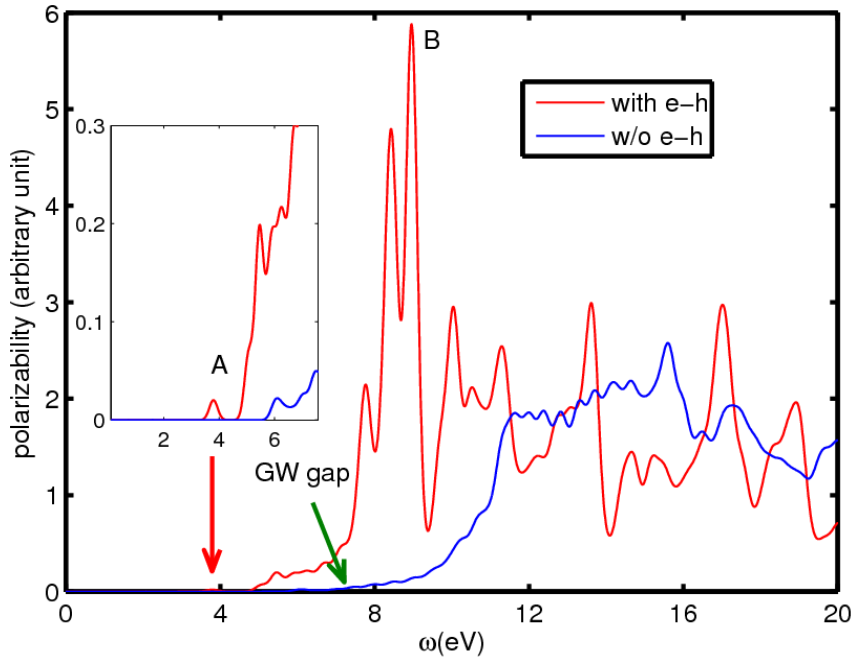

**Figure 4.** Optical absorption spectra of fluorographene with (red line) or without *e-h* interactions (blue line). The green arrow indicates the GW energy gap. The red arrow indicates the position of the first excitonic state, whose optical activity is relatively weak. In order to show the bound exciton in the low-frequency regime, an inset is added.

Through solving the BSE, we are able to calculate the optical response of fluorographene with *e-h* interactions included. In the following calculation, we only consider the optical absorption spectrum with an incident light polarized parallel to the plane of fluorographene. The perpendicularly polarized incident light is not discussed here because of the strong depolarization effect [22]. The optical absorption spectra with and without *e-h* interactions are presented in **Fig. 4**. For the single-particle result (the blue-colored curve), the onset of spectrum locates at the GW band gap (7.3 eV) and the absorption profile increases very slowly until near 10eV, indicating that the lowest optical transition from the valence band maximum (VBM) to the conduction band minimum (CBM) (indicated by the purple arrow in **Fig. 2**) is weak. This is due to the small spatial overlapping of the wave functions of VBM and CBM. On the other hand, the notable optical absorption starting from 10eV is mainly attributed to the transitions from those high valence bands to the *second conduction bands* with energy varying from 8 eV to 11 eV. This can be understood from the fact that the charge distribution of the second lowest conduction band has a significant overlap with that of the first and second highest valence bands (as shown in **Fig. 3 (a)** and **(c)**), giving rise to a pronounced optical absorption.

When *e-h* interaction is taken into account, drastic changes are found in the absorption spectrum of fluorographene, as plotted by the red curve in **Fig. 4**. First, there is a remarkable global red shift of the whole spectrum (~ 4 eV), implying the attracting nature of *e-h* interactions. Our calculation shows that the whole optical absorption spectrum is dominated by resonant excitonic states because all prominent optical absorption is still above the quasiparticle band gap. This is similar to what have been observed in graphane and other nanostructures. In particular, we have identified that the prominent peak (marked as B in **Fig. 4**) is originated from a few strong resonant excitonic states located around 9.0 eV. Moreover, the resonant nature of those prominent excitonic states can be perceived from the fact that it is a mixing of large number of transitions with a similar energy, mainly involving the top two valence bands and the lowest four conduction bands. Two major transitions that contribute to a typical bright resonant excitonic state are depicted by blue arrows in **Fig. 2 (b)**, one from the VBM to the bottom of the second conduction band and the other from the second valence band to the lowest conduction band but occurring away from the $\Gamma$ point.

In the low-frequency optical absorption spectrum, a bound exciton at 3.8 eV is identified with a huge binding energy (3.5 eV), as shown in the inset of **Fig. 4**. This is more than twice larger compared with that of the binding exciton in graphane (1.6 eV) [12]. Although the optical activity of this bound exciton is comparatively weak, it might play a vital role in the photoluminescence because of the large occupation probability associated with the low energy of this excitonic state. Recent experiment has showed a prominent ultraviolet luminescence at 3.8 eV and our calculated results may be of help to understand this measurement.

To illustrate the charge transfer mechanism of excitons in the fluorographene, we have also showed two typical excitonic wave functions in **Fig . 5**. In the two-particle picture, each exciton no longer represents a single interband transition but rather a coherent superposition of pairs of electron and hole states [19, 20]. For the purpose of visualization, we fix the hole at a position slightly beside a fluorine atom and show the relative distribution of the electron for the exciton A and B of **Fig. 4**. Exciton A transfers the electron to a nearest carbon atom and several adjacent fluorine atoms whereas exciton B transfers the electron into fluorographene plane. This is inferable from the previously presented conduction electronic states since these distribution exhibit a very similar pattern. In addition, exciton A exhibits a fast damping character, indicating that the strong binding nature of this exciton while the charge distribution of exciton B has an extended feature due to its resonant nature.

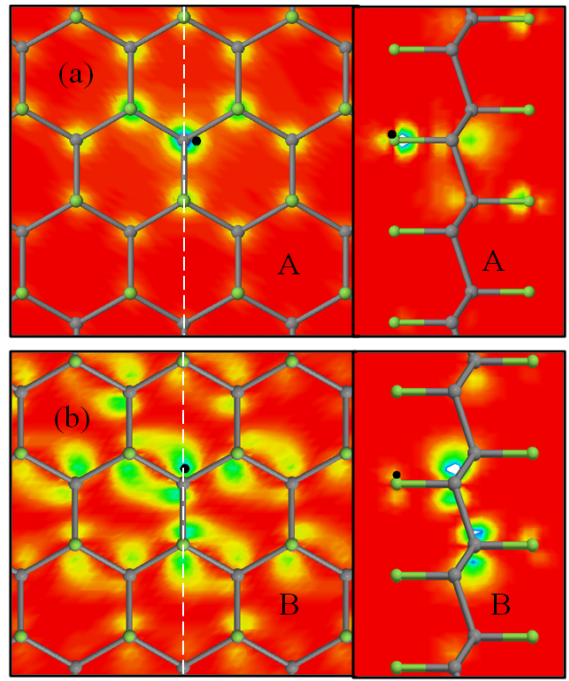

**Figure 5.** Exciton wave functions of fluorographene. The subfigures on the left column are projections of electron distribution $|\chi(\mathbf{r}_h,\mathbf{r}_e)|^2$ for exciton A **(a)** and B **(b)** in the fluorographene plane with a hole position fixed at the black spot. The subfigures on the right column show the corresponding distribution in a plane passing along the armchair direction (white dash line in the left subfigures) and perpendicular to the fluorographene plane.

During preparation of this paper, we become aware of a related theoretical study by D. K. Samarakoon et al. [13] on fluorographene, which has obtained similar trends to those found in the present work.

**CONCLUSIONS**

In conclusion, we have used a first-principles method to reveal the electronic structure of fluorographene with many-electron effects included. Significant self-energy corrections are found to increase the band gap of this material, a 4.3 eV increase. Our results are in consistent with the large band gap observed in experiment. Furthermore, we have unveiled a picture of the optical excitations with an even stronger *e-h* interaction than that in graphane; the whole optical absorption spectrum of fluorographene is dictated by excitonic effects with a huge red shift (approximately 4 eV). Finally, although resonant excitons are the most important factor to decide the optical absorption spectrum of fluorographene, the binding exciton at the low-frequency regime may still play a key role in the photoluminescence measurements.

**ACKNOWLEDGEMENTS**

We acknowledge computational resources support by the Lonestar of teragrid at the Texas Advanced Computing Center (TACC).